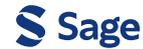

# CerviXpert: A multi-structural convolutional neural network for predicting cervix type and cervical cell abnormalities

Rashik Shahriar Akash[1] , Radiful Islam[1], SM Saiful Islam Badhon[2] and KSM Tozammel Hossain[2]

## Abstract

**Objectives:** Cervical cancer, a leading cause of cancer-related deaths among women globally, has a significantly higher survival rate when diagnosed early. Traditional diagnostic methods like Pap smears and cervical biopsies rely heavily on the skills of cytologists, making the process prone to errors. This study aims to develop CerviXpert, a multi-structural convolutional neural network designed to classify cervix types and detect cervical cell abnormalities efficiently.

**Methods:** We introduced CerviXpert, a computationally efficient convolutional neural network model that classifies cervical cancer using images from the publicly available SiPaKMeD dataset. Our approach emphasizes simplicity, using a limited number of convolutional layers followed by max-pooling and dense layers, trained from scratch. We compared CerviXpert's performance against other state-of-the-art convolutional neural network models, including ResNet50, VGG16, MobileNetV2, and InceptionV3, evaluating them on accuracy, computational efficiency, and robustness using five-fold cross-validation.

**Results:** CerviXpert achieved an accuracy of 98.04% in classifying cervical cell abnormalities into three classes (normal, abnormal, and benign) and 98.60% for five-class cervix type classification, outperforming MobileNetV2 and InceptionV3 in both accuracy and computational demands. It demonstrated comparable results to ResNet50 and VGG16, with significantly reduced computational complexity and resource usage.

**Conclusion:** CerviXpert offers a promising solution for efficient cervical cancer screening and diagnosis, striking a balance between accuracy and computational feasibility. Its streamlined architecture makes it suitable for deployment in resource-constrained environments, potentially improving early detection and management of cervical cancer.

Submission date: 12 June 2024; Acceptance date: 9 October 2024

## Introduction

Cervical cancer, ranked fourth among women globally, arises from cervix tissue, the lower portion of the uterus.[1] An estimated 604,000 new cases and 342,000 deaths occurred for cervix cancer in 2020, with higher mortality rates in underprivileged regions.[2,3] The causes for this cancer include smoking, long-term birth control use,

[1]Department of Computer Science and Engineering, Daffodil International University, Dhaka, Bangladesh
[2]Department of Information Science, University of North Texas, Denton, USA

**Corresponding author:**
Rashik Shahriar Akash, Department of CSE, Daffodil International University, Daffodil Smart City (DSC), Birulia, Savar, Dhaka-1216, Bangladesh.
Email: rashik15-3825@diu.edu.bd

multiple childbirths, multiple sexual partners, and poor menstrual hygiene.[4,5] Early detection is crucial in reducing mortality. The Pap smear test remains the standard screening method, allowing for the identification of precancerous and cancerous cells at an early stage.[6] However, the manual nature of Pap smear analysis is labor-intensive, prone to human error, and highly dependent on the expertise of cytologists.[7]

Pap smear screening involves collecting cells from the cervix's surface and canal using a gentle scraping technique.[8] These collected cells are placed on slides and examined under a microscope to detect abnormalities.[9] However, manual examination poses several challenges, including issues like cell drying, contamination with mucus or blood, and cell clumping, which can obscure the view of individual cells.[10] Automated computer-assisted screening methods, such as AutoPap and FocalPoint, have been developed to address these limitations by capturing microscopic images of the cells. Despite these advancements, these systems still rely on cytologists for the final interpretation, which remains time-consuming and dependent on their expertise.[11]

In response to these limitations, the integration of artificial intelligence (AI)[12] and deep learning into medical imaging has emerged as a transformative approach to improve the accuracy and efficiency of cervical cancer screening.[13] AI-driven tools, particularly those based on deep convolutional neural networks (CNNs),[14] offer the potential to automate the analysis of Pap smear images, thereby reducing the workload on cytologists and minimizing human error.[15,16] These technologies provide a more consistent and rapid means of diagnosing abnormalities by leveraging advanced image analysis capabilities.[17,18] Recent research has explored the use of deep learning models, such as ResNet50, VGG16, MobileNetV2, and InceptionV3, to classify cervical cells with high accuracy and assist in diagnosing cervical cancer.[19,20] However, the computational intensity of these models poses challenges for their widespread implementation, especially in resource-constrained settings where access to high-performance computing may be limited.[21] Addressing this research gap, we propose a deep-learning-based framework—CerviXpert—a computationally inexpensive method for classifying cervix cancer using cell images collected via Pap screening. Our key focus is developing an inexpensive method computationally without compromising diagnostic performance. We essentially develop a multi-structural CNN to solve two problems. The problem asks to classify cell abnormality types, which are normal, abnormal, and benign. These three classes form five different cell types (see Table 1), which the second problem aims to differentiate.

Our key contributions are as follows:

C1. We present a computationally inexpensive multi-structural CNN to identify cervix cancer and cell types with relatively high accuracy.

Table 1. Cervix types and cervical cells.

| Cell | Category | Number of images |
|---|---|---|
| Abnormal | Dyskeratotic | 223 |
| Abnormal | Koilocytotic | 238 |
| Benign | Metaplastic | 271 |
| Normal | Parabasal | 108 |
| Normal | Superficial/intermediate | 126 |
| **Total** | | **966** |

C2. We perform extensive experiments to evaluate our method against existing methods on a real-world dataset.

C3. We explained the complexities between our method and existing methods in detail.

The rest of the article is organized as follows. The "Literature review" section discusses the relevant papers, the "Dataset" section describes the dataset used in the study, the "Methodology" section outlines the methodology employed for the research, the "Results" section presents the findings and results obtained from the analysis, the "Discussion" section discusses the implications and significance of the results. The "Limitations" section acknowledges the limitations of the study and suggests directions for future work. Finally, the "Conclusion" section summarizes the key findings and highlights the overall contributions of the study.

## Literature review

We present pertinent research in this section.

Cervical cancer detection is a new topic compared to the rapid development of computer-aided diagnostics (CAD). In terms of predicting the correct label of a cell being cancerous or non-cancerous, an expert has almost similar accuracy to AI, according to Kurika and Sundado,[22] the Xception architecture, a deep learning model developed using CNNs, 488 photos representing 117 women with cancer and 509 cervix cell images obtained from 181 healthy individuals after giving 50 training epochs. When given a T2-weighted picture, deep learning demonstrated better diagnostic ability for detecting cervical cancer than radiologists.[23–29]

Bhavani and Govardhan[30] showed that support vector machines (SVMs), logistic regression (LR), decision trees (DTs), k-nearest neighbors (KNNs), and random forests (RFs) classification methods have found that the ensemble method produces the best results. DT produces the highest accuracy with over 91.2% accuracy. The second highest was RF with 90.6% accuracy. Similarly, to examine the causes of cervical cancer, this study introduces a DT classification technique. To identify the most useful characteristics for predicting cervical cancer, the authors by Tanimu et al.[31] exhaustively investigated the feature selection methods of recursive feature elimination (RFE). The DT improved results by selecting characteristics from RFE and synthetic minority over-sampling technique, reaching a sensitivity of 100% and an accuracy of 98.72%. The lack of accuracy was caused by missing values due to a numeric data set. In terms of practicality and usefulness, information extracted from visualized images is a little bit ahead.

For instance, according to Urushibara et al.,[32] for the methodology, a vast number of algorithms were introduced, including DT, SVM, KNNs, LR, adaptive boosting, gradient boosting, RF, and XGBoost. Both prediction and classification results from this study have met expectations. SVM was the highest in terms of accuracy with near perfect 99%. The author of Fekri-Ershad and Alsaffar[33] suggests using deep features from the ResNet-34, ResNet-50, and VGG-19 CNNs to feed a multi-layer perceptron neural network. The feature extraction stage is separated from the classification stage in the machine learning technique that underpins the classification stage. The proposed method demonstrated a high accuracy of 99.23% when tested against the Herlev benchmark database. According to the study's findings, the proposed method offers greater accuracy than using the two networks individually and many other existing methods.

The study of Fekri-Ershad and Ramakrishnan[34] suggests a two-stage method for detecting cervical cancer using pap smear images. The first stage involves the extraction of texture information from the cytoplasm and nucleolus using a modified uniform local ternary pattern descriptor. The second stage uses a multi-layer feed-forward neural network that has been optimized for image classification. A genetic algorithm is used to optimize the neural network by determining the ideal number of hidden layers and nodes. On the Herlev database, the proposed approach is assessed and contrasted with existing methods. The findings demonstrate that the suggested method is insensitive to image rotation and has a greater detection accuracy than the methods that were compared.

By suggesting an end-to-end classification of cervical cells using deep features, the author of Rahaman et al.[35] seeks to increase the accuracy of cervical cancer screening. Using two publicly accessible datasets, SIPaKMeD and Herlev, the proposed method is evaluated and contrasted with various basic Deep Learning models and late fusion (LF) techniques. The suggested approach provides end-to-end categorization without relying on hand-crafted characteristics or pre-segmentation. The study of Fahad et al.[36] developed an automated cervical cancer diagnostic system using AI, focusing on segmenting and classifying cervical cancer cell types from histopathology images. They utilized a graph convolutional network (GCN) on a graph dataset constructed from handcrafted features, achieving an impressive accuracy of 96.17% on the SipakMed dataset with 30 features.

Srinivasan et al.[37] utilized associated histogram equalization to enhance image edges, followed by finite ridgelet transform and feature extraction methods such as gray-level run-length matrices, moment invariants, and enhanced local ternary patterns. These features are then classified using a feed-forward neural network to differentiate between normal and abnormal images, achieving segmentation of cancerous regions with morphological operations. The system outperforms traditional methods with 98.11% sensitivity, 98.97% specificity, and 99.19% accuracy, demonstrating superior detection and segmentation capabilities compared to existing techniques. Some also explorers data protection,[38,39] blockchain-based data encryption[40] in medical images. These approaches not only improve diagnostic accuracy but also ensure the security and integrity of sensitive medical data, highlighting the critical role of advanced image processing and encryption techniques in enhancing overall healthcare outcomes.

Chauhan et al.[41] presented a hybrid deep feature concatenated network combined with two-step data augmentation to detect cervical cancer through binary and multiclass classification of Pap smear images. The features extracted from fine-tuned VGG-16, ResNet-152, and DenseNet-169 models pre-trained on ImageNet, the proposed model achieved an accuracy of 97.45% for five-class classifications. The author of Deo et al.[42] introduces CerviFormer, a cross-attention-based transformer model designed for accurate cervical cancer classification in Pap smear images. CerviFormer efficiently handles large-scale input data, offering robust performance on two publicly available datasets: Sipakmed and Herlev, and achieving 93.70% accuracy for three-state classification on Sipakmed and 94.57% for two-state classification on Herlev. This study develops an end-to-end architecture using three pre-trained models and a novel fuzzy rank-based ensemble for cervical cancer prediction in Pap smear images. Integrating advanced augmentation techniques, the model achieves an accuracy of 97.18% and $F1$ score of 97.16% on the SIPaKMeD dataset.

## Datasets

We use a publicly available dataset, SIPaKMeD,[43] to evaluate the existing and proposed methods. We extracted five types of cells, which constituted 966 images. The breakdown of these images is shown in Table 1. These cells are collected via a Pap test (a.k.a Pap smear) and labeled by domain experts. Figure 1 illustrates some samples of cervical cells.

These five types of cells are grouped into three broad categories in terms of cancerous nature. For three-class classification, the dataset was prepared in three categories—normal, abnormal, and benign class. Normal and abnormal classes have 695 images each, while the benign class has 271 images.

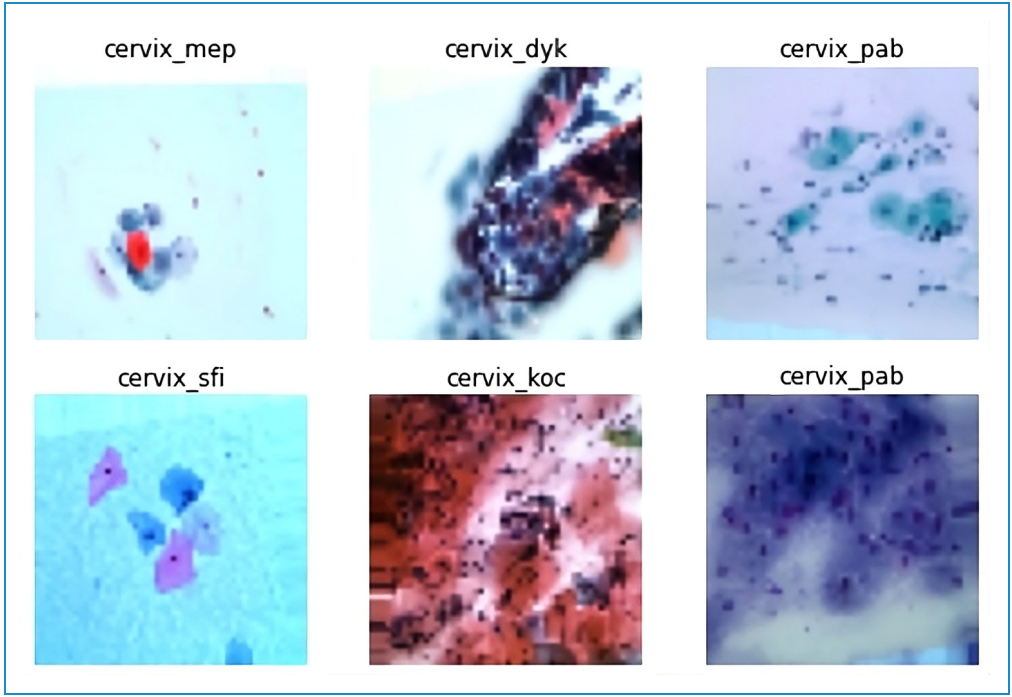

**Figure 1.** Examples of different cervical cells collected during Pap smear tests: (top left to right: metaplastic, dyskeratotic, and parabasal) and (bottom left to right: superficial, koilocytotic, and parabasal).

### Data pre-processing

In our study, we applied a series of image augmentation techniques to enhance the robustness and generalization of the classification models. These techniques were designed to create variations of the training images, thus aiding the models in learning more comprehensive features and preventing overfitting. We began with rotation, applying a 45-degree rotation to the images. This transformation helps the model become invariant to the orientation of cervical cells, ensuring that the model can detect cells regardless of their rotation in the image. To address variations in vertical orientation, we employed vertical flipping of the images. This augmentation allows the model to handle different vertical orientations, which might occur due to diverse image acquisition conditions. Additionally, we utilized zoom to enhance the model's ability to recognize cervical cells at varying scales. By zooming in on the images with a factor of 1.2, we provided the model with examples of cells at different magnifications, improving its performance in handling scale variations.

For more sophisticated augmentation, we applied elastic transformations. This technique introduces non-linear deformations in the images, modeled by the following equation:

$$f'(x, y) = f(x + \mathbf{v}_x(x, y), y + \mathbf{v}_y(x, y)) \quad (1)$$

where $\mathbf{v}_x(x, y)$ and $\mathbf{v}_y(x, y)$ represent the displacement fields generated through Gaussian smoothing with parameters $\sigma$ and $\alpha$. This method helps the model to generalize better by incorporating realistic distortions in the shape of cervical cells. Finally, we used contrast limited adaptive histogram equalization (CLAHE) to enhance image contrast. The CLAHE technique adjusts the histogram of the image adaptively, with the transformation function of the following equation:

$$T(i) = \frac{N_c(i)}{N} \quad (2)$$

where $N_c(i)$ is the cumulative histogram value for the intensity $i$, and $N$ represents the total number of pixels. This method improves contrast, making it easier to detect details in cervical cells across different illumination conditions. Augmented image visualizations can be found in Figure 2. As detailed in Table 2, the number of samples for each class is shown after image augmentation.

### Data setting

A total of 25,000 images of cervical cells have been annotated and included in the SIPaKMeD enhanced dataset. Table 1 describes the number of images from each category and cell. Seventy percent of each class's dataset is utilized for training, 20% for validating, and 10% for testing. Here, a five-cell categorization was done (dyskeratotic, koilocytotic, metaplastic, superficial, and parabasal). Some examples of advanced artificial neural networks include the CNN. The renowned computer scientist developed the CNN while thinking about how the brain works. A ConvNet is built up of many layers. Figure 3 clearly visualize the data setting layers, and Figure 4 visualizes the execution of the system.

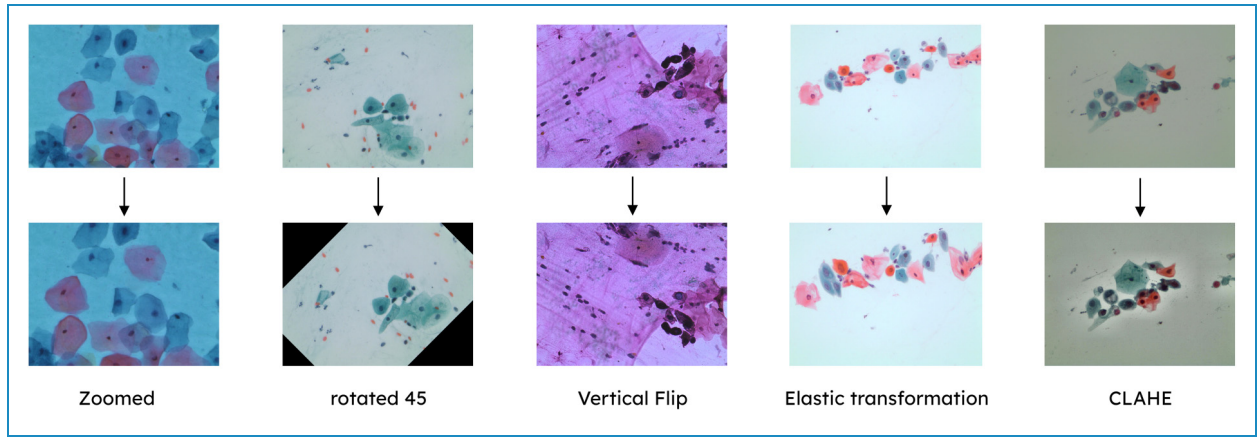

**Figure 2.** Visualization of image augmentation techniques. The figure displays five different augmentation operations applied to some images. The top row shows the original images, while the bottom row illustrates the results after applying each augmentation technique.

**Table 2.** Data statistics after applying image augmentation techniques to create an enhanced sipakmed dataset.

| Cell | Category | Number of images |
| --- | --- | --- |
| Abnormal | Dyskeratotic | 5000 |
| Abnormal | Koilocytotic | 5000 |
| Benign | Metaplastic | 5000 |
| Normal | Parabasal | 5000 |
| Normal | Superficial/intermediate | 5000 |
| **Total** | | **25,000** |

- Forms of layers: First, we wll run a $100 \times -by-100 \times -by-3$ pixel picture through a CNN.
- Input layer: The input layer contains the image's raw data and has the dimensions $100 \times 100 \times 3$.
- Activation function layer: The activation function layer, is also known as the transfer function layer. Activation functions may be split into two categories. To begin, there is the linear activation function and the nonlinear activation function. There are $100 \times 100 \times 12$ pixels in the final layer's output.
- Convolutional layer: The dot product of all filters applied to a given picture patch is computed at the convolution layer. Assuming a $100 \times 100 \times 3$ input picture, the output will not have the same dimensions if the filter number used is 7. It will produce $100 \times 100 \times 7$ as an output.
- Fully connected layer: Unlike partially connected layers, fully linked layers get data from the layer below them. To do this, it flattens the array into a one-dimensional array of the same length as the class count.
- Pooling layer: The pool layer is often placed after the activation function layer. Many distinct pool layer varieties exist. In terms of pool layering methods, max pooling is among the most popular options. The output dimension is $50 \times by50 \times 12$ if it is utilized with a pool size of $2 \times by2$ and a stride of 2.

## Methodology

In this section, we formulate the problem, describe existing methods, and present our approach. This study is experimental in nature and was conducted over a period of 6 months, from January to June 2024, in Bangladesh.

Several methods or models are used for the classification of both three-class and five types of cells. We detail the approach we used in this study and demonstrate the functionality of the suggested classifiers. Each approach leads to a variety in validation loss and accuracy.

In this study, we introduce CerviXpert, a multi-structural CNN, as a method for identifying cervical cancer. CerviXpert is designed to address the computational expense associated with existing methods such as ResNet50, VGG16, MobileNetV2, and InceptionV3, which are based on deep CNNs. CNNs are gaining more and more attention with each passing day. Using a deep CNN is the most effective method for accelerating the process of diagnosing crop problems and delivering the appropriate treatment in a short time. Through these experiments and analyses, we aim to showcase CerviXpert as a promising solution for efficient cervical cancer screening and diagnosis, emphasizing its ability to strike a balance between accuracy and practical feasibility in comparison to existing methods.

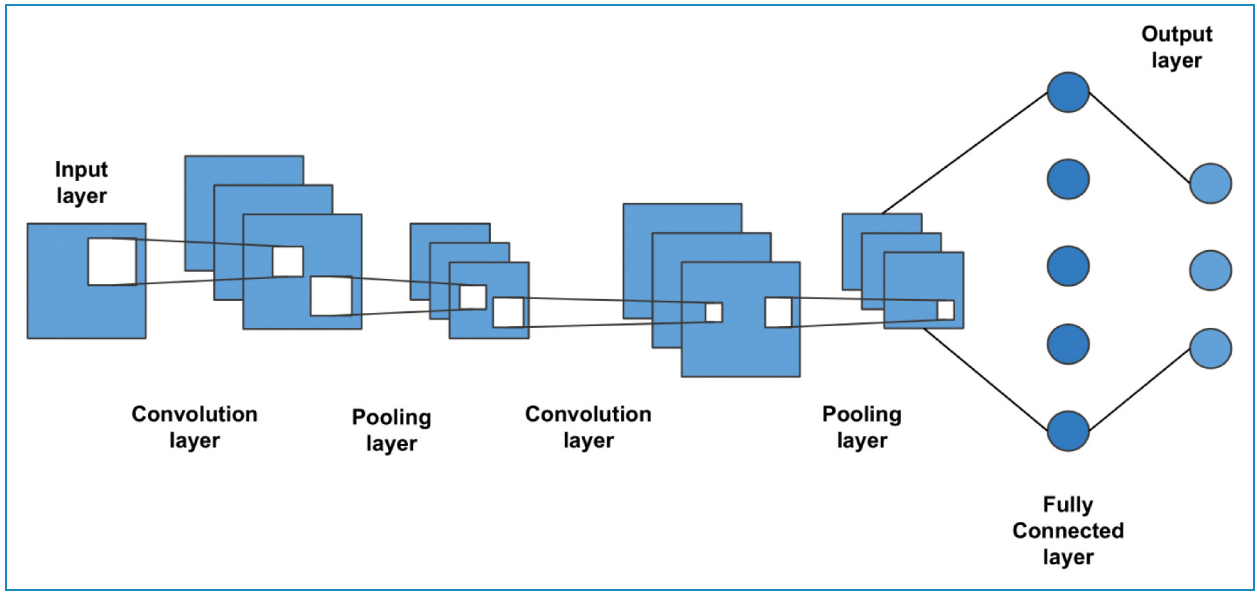

**Figure 3.** Data setting layers.

Now, we describe the existing methods for classifying cervix cancer.

### Existing methods

*Convolutional neural network.* A form of neural network called a convolutional neural network (CNN) is frequently employed in image and video recognition applications. Convolutional layers are used in CNNs to automatically recognize and extract characteristics from photos or videos. The foundation of a CNN is made up primarily of convolutional layers. They take an input image or feature map and apply a series of learnable filters—also referred to as kernels or weights—to produce a set of output feature maps that reflect various facets of the input. The filters apply a dot product between the filter and a tiny section of the input at each place, creating a new feature map that enumerates the filter's existence at each location as the filters move over the input image or feature map.

A CNN comprises various layers designed for effective feature extraction and hierarchical representation learning.[44–46] Convolutional layers apply filters to input data, generating feature maps that capture spatial hierarchies. Pooling layers reduce spatial dimensions through techniques like max-pooling or average-pooling, preserving essential features. Activation layers introduce non-linearity, crucial for complex pattern recognition. Fully connected layers connect every neuron in one layer to all neurons in the next, facilitating the network's ability to categorize input data. Dropout layers mitigate overfitting by randomly excluding neurons during training. Additionally, batch normalization layers enhance training stability and speed by normalizing the output of preceding layers, reducing internal covariate shifts.

The selection of models for cervical cell classification was based on several factors, including architectural characteristics, performance, and resource efficiency.

*ResNet50.* ResNet-50 is a 50-layer deep neural network design including convolutional, batch normalization, and rectified linear unit (ReLU) activation layers in every layer. The structure is broken up into several phases, each of which is made up of various remaining building components. To summarize, each residual block has numerous layers and a fast route link. The input of the block is connected directly to its output through the shortcut connection, which avoids one or more intermediate nodes. In residual connections, the network is trained not on the original mapping between an input and an output block, but on the mapping left behind after the block's processing is complete. As a result, vanishing gradients are no longer an issue, allowing for considerably deeper network training. Bottleneck architecture is used by ResNet-50 to lower the computational cost by decreasing the number of filters in the $1 \times 1$ convolutional layers. The next step is an image feature extraction process using $3 \times 3$ convolutional layers. As a finishing touch, $1 \times 1$ convolutional layers are utilized once more to boost the total number of filters before the output. ResNet-50 has already been trained on the massive ImageNet dataset of over 14 million pictures and 1000 classes. By first pre-training the network, it is able to pick up information from the massive dataset that can be used to additional picture categorization jobs. The performance of the pre-trained network may be enhanced by fine-tuning it on a smaller dataset for a given purpose.

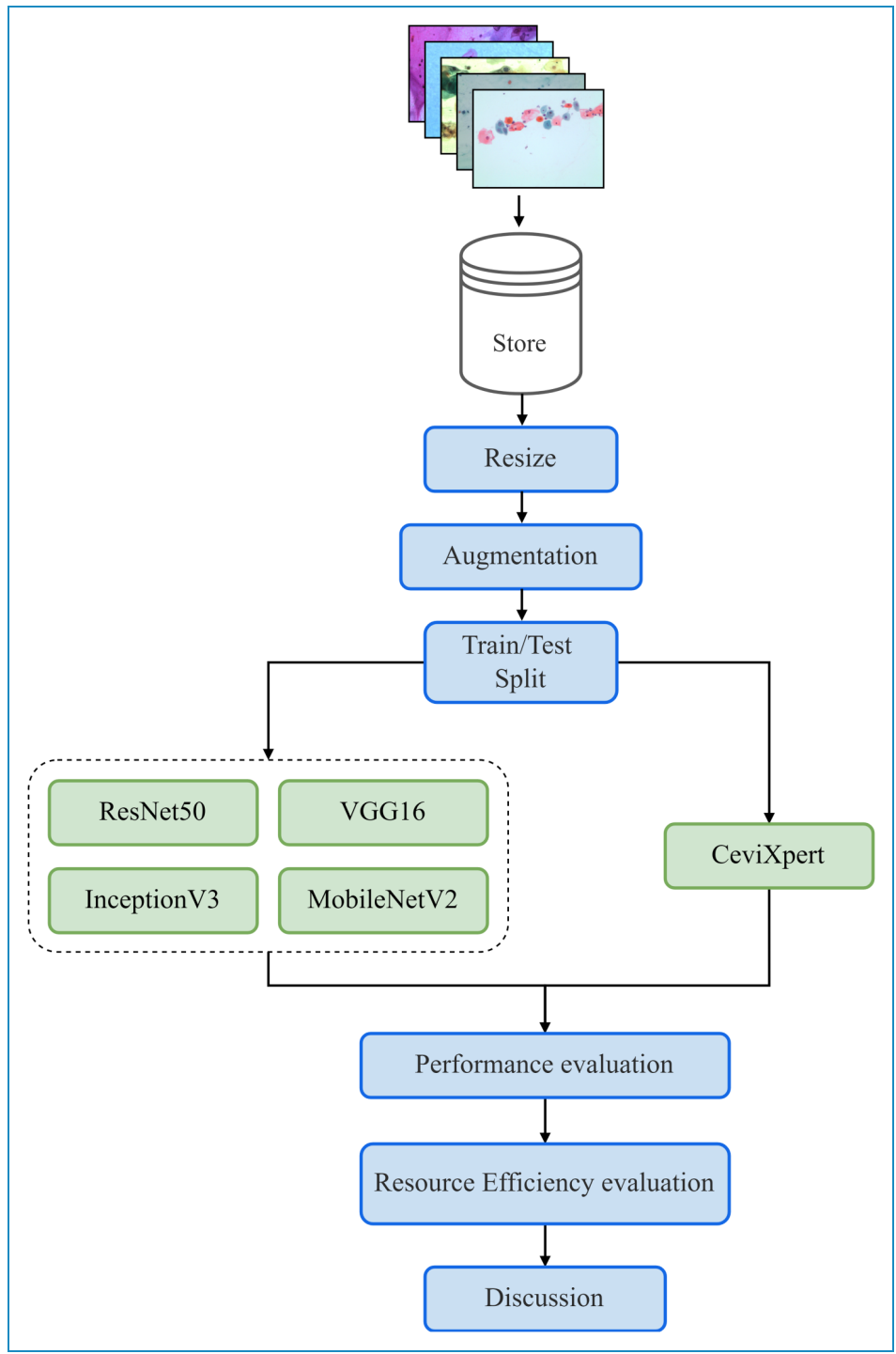

**Figure 4.** Execution of the system.

*VGG16.* VGG16 is a widely recognized deep CNN architecture for the role in image classification tasks. It consists of 16 layers, including 13 convolutional layers and three fully connected layers. A key feature of VGG16 is its use of small $3 \times 3$ convolutional filters, which enable the network to capture fine-grained details in images while maintaining a deep architecture. These convolutional layers are grouped into blocks, each followed by max pooling layers that downsample the spatial dimensions. The network employs ReLU activation functions to introduce non-linearity and batch normalization layers to stabilize and speed up training. By using this deep architecture and compact filters, VGG16 can effectively extract hierarchical features, making it highly effective for image classification and other computer vision tasks.

*InceptionV3.* InceptionV3 is an advanced version of the inception architecture, designed for efficient computation and improved performance in image classification tasks. One of the main innovations in InceptionV3 is its ability to reduce the computational cost while maintaining high accuracy by optimizing the Inception modules. This is achieved by combining multiple convolutional filters, such as $3 \times 3$ and $5 \times 5$ filters, which are responsible for extracting detailed features from images, with $1 \times 1$ convolutional filters that effectively reduce the dimensionality of the data. This dimensionality reduction not only speeds up processing but also decreases the number of parameters in the model, making it more efficient without compromising on accuracy. InceptionV3's streamlined design allows it to capture both local and global features of an image, enhancing its performance across a wide range of computer vision tasks while keeping the computational load manageable.

*MobileNetV2.* MobileNetV2 is a lightweight CNN architecture designed for efficiency, particularly in mobile and embedded devices. Its primary innovation lies in the use of depth-wise separable convolution, which significantly reduces the computational load while maintaining high accuracy. This process involves two key steps: the depth-wise convolution, where a single filter is applied to each input channel independently, and the pointwise convolution, which combines the output of the depth-wise convolution across all channels using $1 \times 1$ convolutions. By separating the standard convolution operation into these two phases, MobileNetV2 drastically reduces the number of parameters and computation time

Figures 5 to 9 are the visualization of used algorithm.

### Proposed method: CerviXpert

In this research, we utilize CerviXpert, a custom CNN model specifically designed for the classification of cervical cell images. The model architecture is tailored to efficiently process and categorize images into either three or five distinct classes, depending on the specific classification task. CerviXpert begins with a series of convolutional layers that extract features from input images of size $100 \times 100$ pixels. The first convolutional layer employs 64 filters to capture low-level features, followed by a max pooling layer to reduce the spatial dimensions of the feature maps. This is succeeded by a second convolutional layer with 128 filters, which detects more complex features, and another max pooling layer to further down-sample the feature maps. The third convolutional layer, equipped with 256 filters, extracts high-level features, and is followed by a final max pooling layer that prepares the feature maps for the subsequent fully connected layers. The extracted features are then flattened into a one-dimensional vector, which is passed through a dense layer with 128 units. This layer aggregates the features and captures intricate relationships between them. The final output layer uses a softmax activation function to provide probability distributions over the class labels, allowing for classification into the predefined number of classes. The model is compiled with the Adam optimizer and SparseCategoricalCrossentropy loss function, suitable for multi-class classification tasks. During training, the model's performance is evaluated based on accuracy. The training process involves adjusting the model's parameters over 10 epochs, using a specified batch size to process data in manageable chunks. The training and validation datasets ensure that CerviXpert generalizes well and performs effectively on unseen data. Figure 9 demonstrates the model architecture. Tables 3 and 4 show the architectural parameter and training parameter of proposed model.

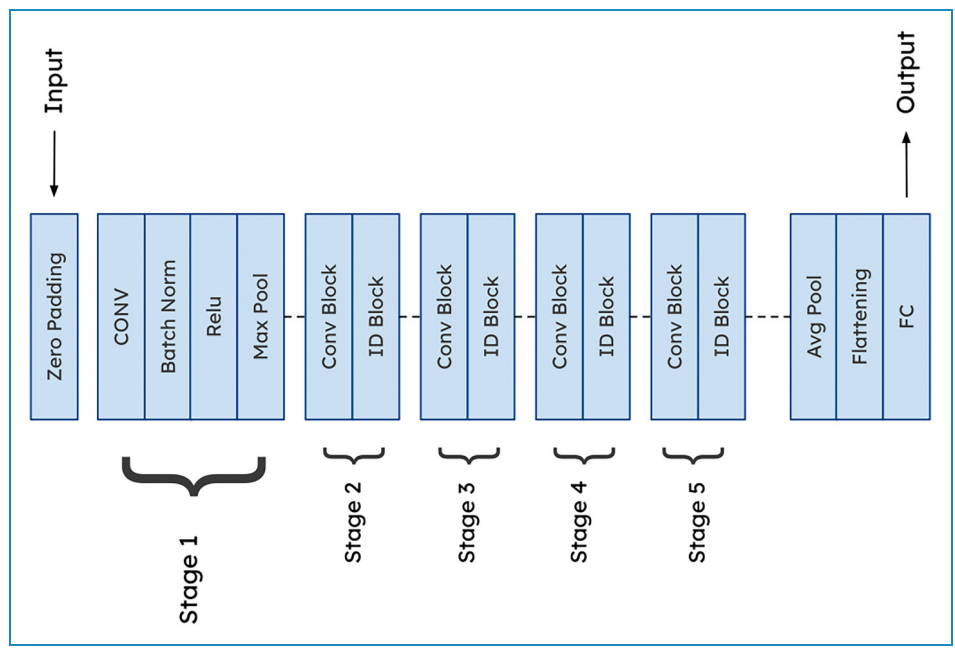

**Figure 5.** ResNet50 architecture.

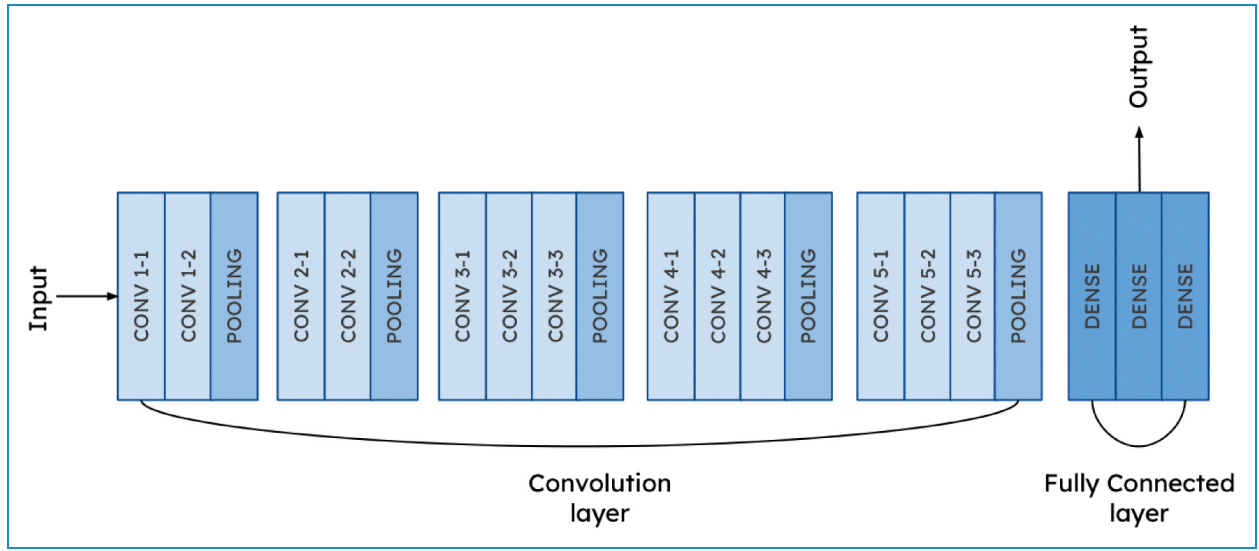

**Figure 6.** VGG16 architecture.

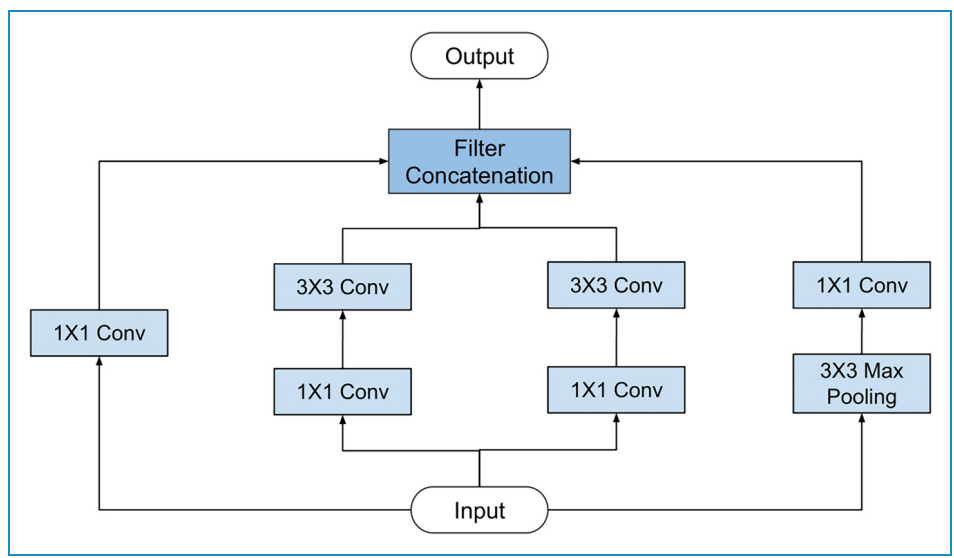

**Figure 7.** InceptionV3 architecture.

*Why CerviXpert?.* The key novelty of our approach lies in the simplicity and efficacy of CerviXpert's CNN architecture. While pre-trained models like ResNet50, VGG16, and InceptionV3 boast complex architectures trained on vast datasets like ImageNet, our model CerviXpert diverges by embracing simplicity. By leveraging a streamlined architecture comprising a few convolutional layers followed by

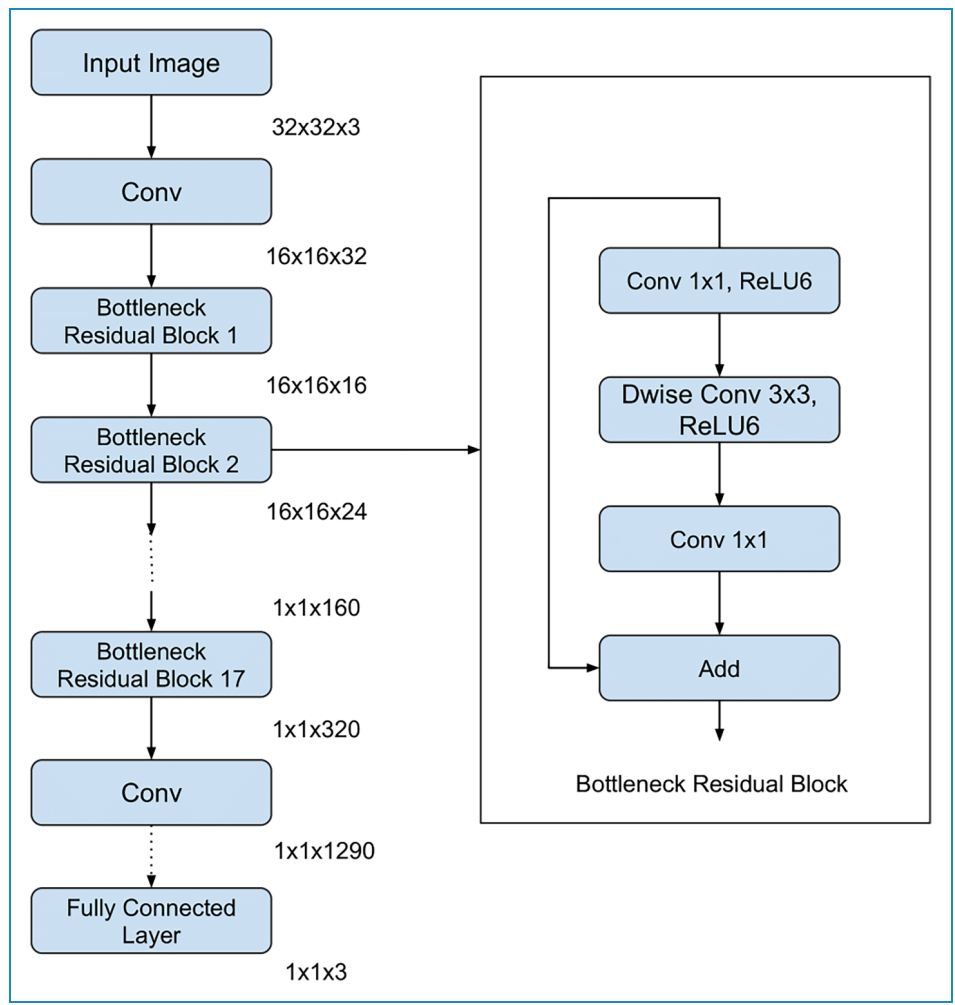

**Figure 8.** MobileNetV2 architecture.

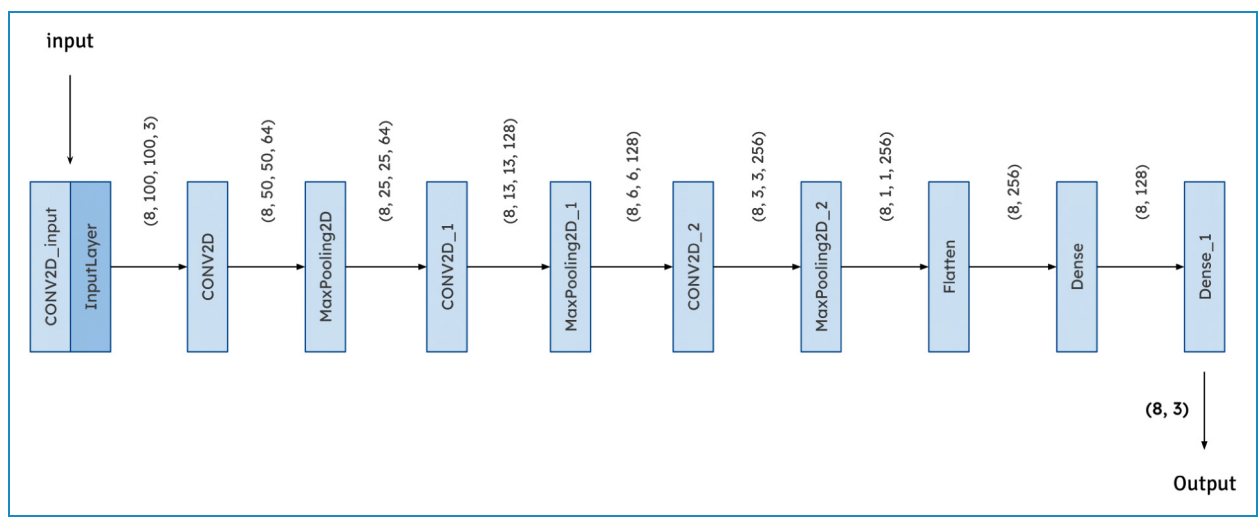

**Figure 9.** CerviXpert architecture.

**Table 3.** Architectural parameter of CerviXpert.

| Layer type | Number of filters/units | Kernel size | Stride | Activation function |
|---|---|---|---|---|
| Conv2D | 64 | (3, 3) | (2, 2) | ReLU |
| MaxPool2D | – | (2, 2) | – | – |
| Conv2D | 128 | (3, 3) | (2, 2) | ReLU |
| MaxPool2D | – | (2, 2) | – | – |
| Conv2D | 256 | (3, 3) | (2, 2) | ReLU |
| MaxPool2D | – | (2, 2) | – | – |
| Dense | 128 | – | – | ReLU |
| Dense | 3 | – | – | Softmax |

ReLU: rectified linear unit.

**Table 4.** Training parameter of CerviXpert.

| Training parameter | Value |
|---|---|
| Optimizer | Adam |
| Loss function | SparseCategoricalCrossentropy |
| Metrics | Accuracy |
| Epochs | 25 |

max-pooling and dense layers, our model exhibits remarkable efficiency in both training and inference. Furthermore, CerviXpert is trained from scratch, eschewing reliance on pre-existing features learned from unrelated datasets. This departure from transfer learning underscores our commitment to tailoring the model specifically for the nuances of cervical cell classification. Despite starting with randomly initialized weights, our model achieves a commendable accuracy of 98.60%, showcasing its ability to discern relevant features directly from the dataset. The superiority of CerviXpert is further underscored by its parameter efficiency. With fewer parameters compared to pre-trained models, our custom architecture not only conserves memory but also accelerates inference speed, making it an appealing choice for resource-constrained environments.

### *Model evaluation*

The primary focus of this work is to provide a method for reliable cervical cancer diagnosis. Pap smear categorization is done in this regard. The primary metric for assessing the success of such situations is the degree to which they can be classified correctly. As per "Data pre-processing" section, samples from the SIPaKMeD dataset can be divided into three or five separate categories. For classifications, the accuracy, precision, and recall may be determined using equations (3) to (5).

$$\text{Accuracy} = \frac{TP + TN}{TP + TN + FP + FN} \quad (3)$$

where true positives (TP), false positives (FP), true negatives (TN), and false negatives (FN).

There is a significant difference in the likelihood of making a false positive or negative diagnosis between the two groups, making cervical cancer diagnosis a binary classification challenge. In other words, the chance of incorrectly identifying a healthy individual as having cervical cancer is substantially lower than the risk of incorrectly diagnosing an infected individual as being healthy. The effectiveness of the suggested method is measured more by its precision and recall than by its accuracy in this context. The following equations to get these measures of performance:

$$\text{Precision} = \frac{TP}{TP + FP} \quad (4)$$

$$\text{Recall} = \frac{TP}{TP + FN} \quad (5)$$

We have clearly described the execution of the full system in Figure 4.

### **Results**

To show the efficacy of CerviXpert, we compare the method against existing methods (see "Results" section) using a real-world dataset (see "Datase" section). In particular, we address the research questions:

**RQ1.** How does the proposed model fare against the existing models regarding traditional performance measures? (see "Predictive performance of the model"section).

**RQ2.** How does the model perform in terms of computing time and complexity? (see "Computing performance of the model" section).

**RQ3.** Does the model show robustness? (see "Robustness of the model" section).

### *Experimental setup*

We use a publicly available SIPaKMeD dataset for evaluating the methods. This augmented dataset contains 25,000 images from five types of cervical cells and falls into three broad classes—normal, abnormal, and benign—from the perspective of abnormality. Given the images, we devise two prediction tasks: (a) identify the cervix type and (b) determine the cell abnormality. Each of the five cell types has ≈5000 instances. As for the abnormality task prediction, the number of normal, abnormal, and benign instances are 10,000, 10,000, and 5000, respectively.

### Predictive performance of the model (RQ1)

The study evaluates the performance of four pre-trained deep learning models (InceptionV3, Resnet50, Vgg16, and MobilenetV2) against the proposed method, CerviXpert, using the dataset. In this setting, 70% of the data is used to train the model, 20% of the data is used to validate the model, and the final 10% of the data is used to assess how well the trained model performed. The results showed that the Resnet50 model achieved the highest accuracy of 99.55% in the three-class cervical cell abnormalities prediction task, followed by Vgg16 with 99.50% accuracy, respectively. CerviXpert, MobileNetV2, and InceptionV3 achieved an accuracy of 98.04%, 86.95%, and 75.15%, and in the five-class cervix type prediction task, the Resnet50 model again achieved the highest accuracy of 99.56%, followed by Vgg16, CerviXpert, and MobilenetV2 with 99.48%, 98.60%, and 82.79% accuracy, respectively. InceptionV3 achieved an accuracy of 62.42%. CerviXpert stands out by surpassing both MobileNetV2 and InceptionV3 models in terms of both accuracy and computational efficiency. While achieving superior accuracy compared to these models, CerviXpert also outperforms them in computational demands.

Table 5 summarizes the performance of existing deep learning methods and the literature's works on the same topic.

By examining the CerviXpert model's accuracy, precision, recall, and $F1$ score, the ultimate performance is evaluated. The results of these five models for five and three classes are shown in Table 6.

### Computing performance of the model (RQ2)

Tables 7 and 8 summarize the resource utilization during training and testing. The hardware configuration comprised a system equipped with two logical CPUs and a total RAM capacity of 12.67 gigabytes (GB). Leveraging the computational prowess of a Tesla T4 GPU, with a dedicated graphics processing unit (GPU) memory total of 15,360 megabytes (MB). This environment is provided by Google Colaboratory.

The exploration of tradeoffs between accuracy and resource efficiency among models is crucial for identifying the most suitable model for a given task. In our research, we analyzed various resources utilized during both the training and testing phases, including time, RAM usage, GPU usage, and model size. Here's a breakdown of the tradeoffs observed:

*Training time.* The training time of a model significantly impacts its efficiency and resource utilization during the training phase. We observed variations in training times among the models, with ResNet50 and VGG16 having longer training times compared to MobileNetV2 and our custom model (CerviXpert). The longer training times of ResNet50 and VGG16 are attributed to their deeper

Table 5. An overview of the comparison between CerviXpert and other advanced methods for cervical cancer detection.

| Paper | Year | Model | Dataset | Accuracy |
|---|---|---|---|---|
| Fang et al.[47] | 2024 | DIFF | SIPaKMeD | 96.02% |
| Kalbhor et al.[48] | 2023 | ResNet50 (fine tuned) | SIPaKMeD | 95.33% |
| Attallah[49] | 2023 | CerCan-Net | SIPaKMeD | 97.7% |
| Chen et al.[50] | 2023 | MSCCNet | SIPaKMeD | 97.90% |
| Pramanik et al.[51] | 2023 | MSENet | SIPaKMeD | 97.21% |
| Maurya et al.[52] | 2023 | Transformer + CNN | SIPaKMeD | 97.6% |
| Yaman and Tuncer[53] | 2022 | Cubic SVM | SIPaKMeD | 98.26% |
| Mousser et al.[54] | 2022 | Incremental deep tree | SIPaKMeD | 93.00% |
| **Ours** | **2024** | **CerviXpert** | **SIPaKMeD** | **98.60%** |

CNN: convolutional neural network; SVM: support vector machine.

**Table 6.** Performance of models for five and three classes. Third, fourth, and fifth columns represent the precision, recall, and *F*1 score for five class classification and seventh, eighth, and ninth columns represent the precision, recall, and *F*1 score for three class classification.

| Model | Five class accuracy | Precision | Recall | F1 score | Three class accuracy | Precision | Recall | F1 score |
|---|---|---|---|---|---|---|---|---|
| **ResNet50** | **99.56%** | **99.57%** | **99.53%** | **99.52%** | **99.55%** | **99.50%** | **99.52%** | **99.50%** |
| VGG16 | 99.48% | 99.47% | 99.47% | 99.42% | 99.50% | 99.47% | 99.47% | 99.44% |
| MobileNetV2 | 82.79% | 85.98% | 86% | 85.81% | 86.95% | 84.60% | 82.75% | 83.04% |
| InceptionV3 | 62.42% | 74.2% | 71.9% | 71.5% | 75.15% | 66.78% | 63.31% | 60.05% |
| CerviXpert | 98.60% | 98.46% | 98.17% | 98.38% | 98.04% | 98.05% | 98.01% | 98.00% |

**Table 7.** Computational performance of CerviXpert on training data. Here *S* denotes second and MB denotes mega byte.

| Model | Training time | Memory usage | Total params | Trainable params | Trained model size |
|---|---|---|---|---|---|
| InceptionV3 | 349.60 S | 16.97 MB | 22065443 (84.17 MB) | 262659 (1.00 MB) | 87.1 MB |
| ResNet50 | 360.81 S | 99.85 MB | 23850371 (90.98 MB) | 262659 (1.00 MB) | 93.5 MB |
| VGG16 | 376.84 S | 14.72 MB | 14780739 (56.38 MB) | 66051 (0.26 MB) | 57.0 MB |
| MobileNetV2 | 193.36 S | 16.66 MB | 2422339 (9.24 MB) | 164355 (0.63 MB) | 11.0 MB |
| **CerviXpert** | **100.17 S** | **15.45 MB** | **404099 (1.54 MB)** | **404099 (1.54 MB)** | **4.7 MB** |

**Table 8.** Computational performance of CerviXpert on testing data.

| Model | Testing time (second) | Avg GPU usage |
|---|---|---|
| InceptionV3 | 1.90 | 65.53% |
| ResNet50 | 1.55 | 37.10% |
| VGG16 | 0.68 | 78.50% |
| MobileNetV2 | 0.98 | 32.00% |
| **CerviXpert** | **0.21** | **8.03%** |

architectures and higher computational complexity. However, these models also tend to achieve higher accuracy due to their ability to capture complex features from the data.

*RAM usage.* RAM (random access memory) usage during training affects the scalability and performance of the training process. InceptionV3 and ResNet50 exhibited higher RAM usage compared to other models, consuming more memory resources. The higher RAM usage of these models is attributed to their larger model sizes and deeper architectures. While high RAM usage may enable the models to process larger amounts of data, it can also lead to memory constraints and performance bottlenecks, especially when dealing with large datasets.

*GPU usage.* GPU (graphics processing unit) utilization is crucial for accelerating deep learning tasks, particularly during model training and inference. VGG16 demonstrated higher GPU utilization during testing compared to other models, indicating more intensive computational processing. Conversely, our custom model, CerviXpert, exhibited lower GPU usage while maintaining competitive accuracy. Efficient GPU utilization is essential for maximizing computational throughput and reducing inference latency, especially in real-time applications.

*Model size.* The size of the trained model directly impacts deployment considerations, storage requirements, and transfer times. Larger models like InceptionV3 and ResNet50 had higher model sizes compared to smaller models like MobileNetV2 and our custom model. While larger models may achieve higher accuracy by capturing more complex patterns, they also require more storage space and computational resources. Smaller model sizes are desirable for deployment in resource-constrained environments, where memory and processing power are limited.

We analyzed the performance of the models in terms of training and testing resource usage, as well as their accuracy.

*Architectural complexity.* Models with deeper architectures and more parameters, such as ResNet50 and VGG16, tend to outperform simpler models like MobileNetV2 and InceptionV3. They can capture more complex patterns and relationships in the data, leading to higher accuracy.

*Resource utilization.* While complex models may offer higher accuracy, they also require more computational resources (e.g. training time, memory, and GPU usage). Simpler models may be more resource-efficient but may sacrifice some accuracy. Achieving the right balance between model complexity and resource efficiency is crucial for optimizing performance.

*Domain specificity.* The effectiveness of a model also depends on the specific characteristics of the dataset and the complexity of the classification task. These models have been extensively studied and validated across various domains, making them ideal choices for comparison with our custom model, CerviXpert. Leveraging well-established architectures allows for a meaningful benchmarking of our custom model's performance against state-of-the-art solutions.

### Robustness of the model (RQ3)

To assess the performance of the pre-trained model along with CerviXpert, we employed five-fold cross-validation. Initially, the dataset was randomly divided into five equally sized subsets. Each subset represents a distinct fold in the cross-validation process. Subsequently, the model underwent training and evaluation iteratively for five rounds, corresponding to the five-fold structure. During each iteration, the model was trained on four of the subsets, while the remaining one subset was designated for validation. This partitioning ensured that the model was exposed to diverse data samples across multiple training and evaluation cycles. This cross-validation approach helps to provide a robust and reliable assessment of the model's performance, as it evaluates the model's generalization capabilities on diverse test sets. The high and consistent classification accuracies obtained across the five folds demonstrate the effectiveness of the proposed CerviXpert model in predicting cervix type and cervical cell abnormalities. The obtained result from each model shown in Table 9.

Figures 10 and 11 show a graph of the CerviXpert model's training and validation accuracy for both three and five classes. The blue line represents the training accuracy, whereas the orange line represents validation accuracy. On the other hand, the training loss is indicated with a blue line and the orange line indicates the validation loss.

**Table 9.** Computational performance of CerviXpert on testing data.

| Model | Accuracy (five fold) |
| --- | --- |
| CerviXpert | 96.79% |
| Resnet50 | 97.76% |
| VGG16 | 97.40% |
| MobileNetV2 | 82.69% |

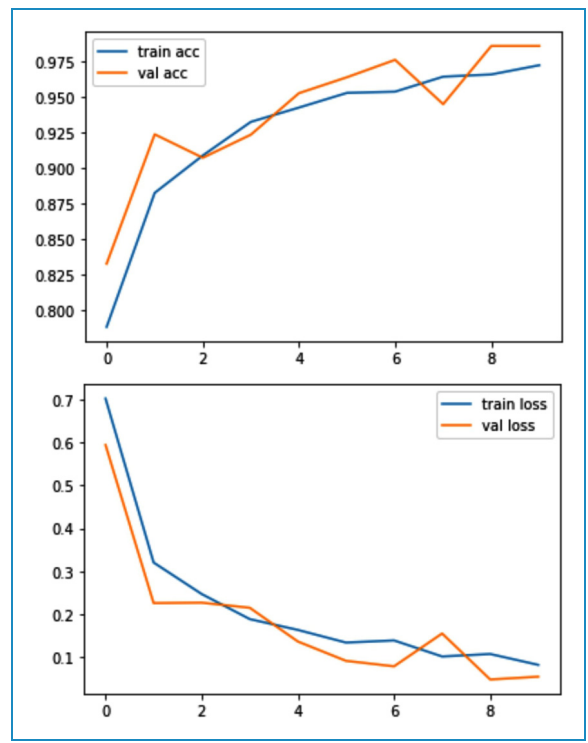

**Figure 10.** Training and validation accuracy and losses for three class in CerviXpert.

### Discussion

Our research underscores the significance of developing resource-efficient deep-learning models for cervical cell classification tasks. CerviXpert, a multi-structural CNN, achieves high diagnostic accuracy with reduced computational demands, making it ideal for resource-constrained environments. Compared to established models like ResNet50, VGG16, MobileNetV2, and InceptionV3, CerviXpert achieved 98.60% accuracy in five-class classification and 98.04% in three-class cervix type classification, demonstrating a superior balance between accuracy and computational efficiency.

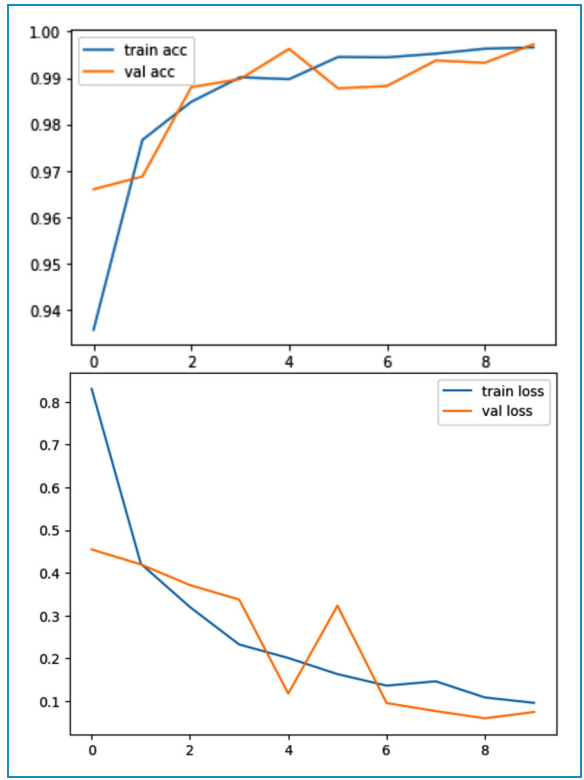

**Figure 11.** Training and validation accuracy and losses for five class CerviXpert.

CerviXpert's efficiency and robustness were validated through extensive testing, including five-fold cross-validation, confirming consistent performance across diverse data subsets. Unlike complex pre-trained models, CerviXpert is trained from scratch, specifically tailored for cervical cell classification without relying on features from unrelated datasets. This approach ensures that the model is not only accurate but also adaptable to new and unseen data, enhancing its utility in clinical settings where rapid and reliable diagnostics are crucial.

The implications of CerviXpert extend to both clinical practice and future research. Clinically, it can assist cytologists by automating the analysis of cervical cell images, potentially reducing diagnostic errors and improving early detection rates. For researchers, CerviXpert serves as a model for developing other resource-efficient deep-learning solutions in medical imaging. Overall, CerviXpert represents a significant step forward in making advanced cervical cancer diagnostics accessible and efficient.

## Limitations and future work

One possible limitation of employing deep CNNs for medical diagnostics is the lack of interpretability of the models. It can be difficult to grasp how the model is making its predictions, which might make it challenging to trust the model in a therapeutic environment. To overcome this, researchers can utilize techniques such as visualization and feature attribution to determine the features the model is using to produce its predictions. Additionally, it is crucial to ensure that the model is resistant to perturbations in the data and generalizes effectively to unseen data. This can be achieved by meticulous validation and testing of the model. In future work, our research aims to advance the field by developing a layered combination of deep learning and machine learning models. The primary objective is to improve the model's performance on datasets with higher dimensions, allowing for a more comprehensive analysis of cervical cell samples.

## Conclusion

In conclusion, this article presents CerviXpert, a novel multi-structural CNN that offers an efficient solution for classifying cervical cell abnormalities using the SiPaKMeD dataset. By focusing on computational efficiency and maintaining high diagnostic accuracy, CerviXpert addresses the limitations of existing deep learning models, which often require significant resources. The model strikes a balance between performance and feasibility, making it well-suited for real-world applications, particularly in resource-limited settings. CerviXpert showcases its ability to effectively classify cervical cell types without the computational overhead of more complex architectures. Its streamlined design enables faster processing times and reduced memory usage while maintaining accuracy, making it a promising tool for automated cervical cancer screening. This approach can potentially ease the burden on healthcare systems, providing a scalable solution for early detection of cervical abnormalities.

**Acknowledgements**: We acknowledge the authors/creators of the SipaKmeD dataset for providing the data.

**Contributorship:** Rashik Shahriar Akash and Radiful Islam: conceptualization; Rashik Shahriar Akash, Radiful Islam, and SM Saiful Islam Badhon: methodology; SM Saiful Islam Badhon: software; KSM Tozammel Hossain: validation; Rashik Shahriar Akash and KSM Tozammel Hossain: formal analysis; Rashik Shahriar Akash, Radiful Islam, and SM Saiful Islam Badhon: investigation; SM Saiful Islam Badhon: resources; Rashik Shahriar Akash and Radiful Islam: data curation; Rashik Shahriar Akash: writing–original draft; all authors: writing–review and editing; Radiful Islam: visualization; KSM Tozammel Hossain: supervision; Rashik Shahriar Akash: project administration.

**Data availability:** The SipaKmeD dataset used for this study is publicly available and can be accessed at https://doi.org/10.1109/ICIP.2018.8451588.

**Declaration of conflicting interests:** The author(s) declared no potential conflicts of interest with respect to the research, authorship, and/or publication of this article.

**Ethical approval:** Ethical approval was not required for this study as it involved the use of a publicly available dataset, SipaKmeD, which contains anonymized data.

**Funding:** The author(s) received no financial support for the research, authorship, and/or publication of this article.

**Guarantor:** Dr KSM Tozammel Hossain acts as the guarantor for the content of this article.

**Informed consent:** N/A as the dataset is publicly available.

**ORCID iD:** Rashik Shahriar Akash  https://orcid.org/0009-0005-7836-2022

## References

1. Girdonia M, Garg R, Jeyabashkharan P, et al. Cervical cancer prediction using Naïve Bayes classification.
2. Setayesh T, Kundi M, Nersesyan A, et al. Use of micronucleus assays for the prediction and detection of cervical cancer: a meta-analysis. *Carcinogenesis* 2020; 41: 1318–1328.
3. Liang Y, Pan C, Sun W, et al. Global context-aware cervical cell detection with soft scale anchor matching. *Comput Methods Programs Biomed* 2021; 204: 106061.
4. Wentzensen N, Lahrmann B, Clarke MA, et al. Accuracy and efficiency of deep-learning-based automation of dual stain cytology in cervical cancer screening. *JNCI J Natl Cancer Inst* 2021; 113: 72–79.
5. Lilhore UK, Poongodi M, Kaur A, et al. Hybrid model for detection of cervical cancer using causal analysis and machine learning techniques. *Comput Math Methods Med* 2022; 2022: 4688327.
6. Al Mudawi N and Alazeb A. A model for predicting cervical cancer using machine learning algorithms. *Sensors* 2022; 22: 4132.
7. Rahaman MM, Li C, Wu X, et al. A survey for cervical cytopathology image analysis using deep learning. *IEEE Access* 2020; 8: 61687–61710.
8. Kaushik K, Bhardwaj A, Bharany S, et al. A machine learning-based framework for the prediction of cervical cancer risk in women. *Sustainability* 2022; 14: 11947.
9. Singh SK and Goyal A. Performance analysis of machine learning algorithms for cervical cancer detection. In: *Research anthology on medical informatics in breast and cervical cancer*, 2023. pp.347–370. IGI Global.
10. Kruczkowski M, Drabik-Kruczkowska A, Marciniak A, et al. Predictions of cervical cancer identification by photonic method combined with machine learning. *Sci Rep* 2022; 12: 3762.
11. Chauhan NK and Singh K. Performance assessment of machine learning classifiers using selective feature approaches for cervical cancer detection. *Wireless Pers Commun* 2022; 124: 2335–2366.
12. Boon SS, Luk HY, Xiao C, et al. Review of the standard and advanced screening, staging systems and treatment modalities for cervical cancer. *Cancers* 2022; 14: 2913.
13. Rose PG, Baker S, Fournier L, et al. Serum squamous cell carcinoma antigen levels in invasive cervical cancer: prediction of response and recurrence. *Am J Obstet Gynecol* 1993; 168: 942–946.
14. Zhang L, Lu L, Nogues I, et al. DeepPap: deep convolutional networks for cervical cell classification. *IEEE J Biomed Health Inform* 2017; 21: 1633–1643.
15. Teixeira JC, Vale DB, Campos CS, et al. Organization of cervical cancer screening with DNA–HPV testing impact on early-stage cancer detection: a population-based demonstration study in a Brazilian city. *Lancet Reg Health Am* 2022; 5: 100084.
16. Ratul IJ, Al-Monsur A, Tabassum B, et al. Early risk prediction of cervical cancer: a machine learning approach. In: *2022 19th international conference on electrical engineering/electronics, computer, telecommunications and information technology (ECTI-CON)*, pp.1–4. IEEE.
17. Fernandes K, Chicco D, Cardoso JS, et al. Supervised deep learning embeddings for the prediction of cervical cancer diagnosis. *PeerJ Comput Sci* 2018; 4: e154.
18. Khamparia A, Gupta D, Rodrigues JJ, et al. DCAVN: cervical cancer prediction and classification using deep convolutional and variational autoencoder network. *Multimed Tools Appl* 2021; 80: 30399–30415.
19. Chen H, Liu J, Wen QM, et al. CytoBrain: cervical cancer screening system based on deep learning technology. *J Comput Sci Technol* 2021; 36: 347–360.
20. Liu W, Li C, Xu N, et al. CVM-Cervix: a hybrid cervical Pap-smear image classification framework using CNN, visual transformer and multilayer perceptron. *Pattern Recognit* 2022; 130: 108829.
21. Raza A, Ayub H, Khan JA, et al. A hybrid deep learning-based approach for brain tumor classification. *Electronics* 2022; 11: 1146.
22. Kurika AE and Sundado TS. Predicting factors of vehicle traffic accidents using machine learning algorithms: in the case of Wolaita zone. *Int J Sci Res Comput Sci Eng* 2020; 8: 105–115.
23. Sahoo P, Saha S, Mondal S, et al. Enhancing computer-aided cervical cancer detection using a novel fuzzy rank-based fusion. *IEEE Access* 2023; 11: 145281–145294.
24. Shi J, Wang R, Zheng Y, et al. Cervical cell classification with graph convolutional network. *Comput Methods Programs Biomed* 2021; 198: 105807.
25. Wadekar S and Singh DK. A modified convolutional neural network framework for categorizing lung cell histopathological image based on residual network. *Healthcare Anal* 2023; 4: 100224.
26. Xue D, Zhou X, Li C, et al. An application of transfer learning and ensemble learning techniques for cervical histopathology image classification. *IEEE Access* 2020; 8: 104603–104618.
27. Tseng CJ and Tang C. An optimized XGBoost technique for accurate brain tumor detection using feature selection and image segmentation. *Healthcare Anal* 2023; 4: 100217.
28. Dweekat OY and Lam SS. Cervical cancer diagnosis using an integrated system of principal component analysis, genetic algorithm, and multilayer perceptron. In: *Healthcare*, volume 10, p.2002. MDPI.
29. Singh J and Sharma S. Prediction of cervical cancer using machine learning techniques. *Int J Appl Eng Res* 2019; 14: 2570–2577.
30. Bhavani C and Govardhan A. Cervical cancer prediction using stacked ensemble algorithm with SMOTE and RFERF. *Mater Today Proc* 2023; 80: 3451–3457.

31. Tanimu JJ, Hamada M, Hassan M, et al. A machine learning method for classification of cervical cancer. *Electronics* 2022; 11: 463.
32. Urushibara A, Saida T, Mori K, et al. Diagnosing uterine cervical cancer on a single T2-weighted image: comparison between deep learning versus radiologists. *Eur J Radiol* 2021; 135: 109471.
33. Fekri-Ershad S and Alsaffar MF. Developing a tuned three-layer perceptron fed with trained deep convolutional neural networks for cervical cancer diagnosis. *Diagnostics* 2023; 13: 686.
34. Fekri-Ershad S and Ramakrishnan S. Cervical cancer diagnosis based on modified uniform local ternary patterns and feed forward multilayer network optimized by genetic algorithm. *Comput Biol Med* 2022; 144: 105392.
35. Rahaman MM, Li C, Yao Y, et al. DeepCervix: a deep learning-based framework for the classification of cervical cells using hybrid deep feature fusion techniques. *Comput Biol Med* 2021; 136: 104649.
36. Fahad NM, Azam S, Montaha S, et al. Enhancing cervical cancer diagnosis with graph convolution network: AI-powered segmentation, feature analysis, and classification for early detection. *Multimed Tools Appl* 2024; 83: 1–25.
37. Srinivasan S, Raju ABK, Mathivanan SK, et al. Local-ternary-pattern-based associated histogram equalization technique for cervical cancer detection. *Diagnostics* 2023; 13: 548.
38. Raghunandan K, Dodmane R, Bhavya K, et al. Chaotic-map based encryption for 3D point and 3D mesh fog data in edge computing. *IEEE Access* 2022; 11: 3545–3554.
39. Sahu M, Padhy N, Gantayat SS, et al. Shadow image based reversible data hiding using addition and subtraction logic on the LSB planes. *Sens Imaging* 2021; 22: 7.
40. Kallapu B, Dodmane R, Thota S, et al. Enhancing cloud communication security: a blockchain-powered framework with attribute-aware encryption. *Electronics* 2023; 12: 3890.
41. Chauhan NK, Singh K, Kumar A, et al. HDFCN: a robust hybrid deep network based on feature concatenation for cervical cancer diagnosis on WSI Pap smear slides. *BioMed Res Int* 2023; 2023: 4214817.
42. Deo BS, Pal M, Panigarhi PK, et al. CerviFormer: a Pap-smear based cervical cancer classification method using cross attention and latent transformer. *arXiv preprint arXiv:230310222*, 2023.
43. Plissiti ME, Dimitrakopoulos P, Sfikas G, et al. Sipakmed: a new dataset for feature and image based classification of normal and pathological cervical cells in pap smear images. In: *2018 25th IEEE international conference on image processing (ICIP)*, pp.3144–3148. IEEE.
44. Manna A, Kundu R, Kaplun D, et al. A fuzzy rank-based ensemble of CNN models for classification of cervical cytology. *Sci Rep* 2021; 11: 14538.
45. Chen W, Shen W, Gao L, et al. Hybrid loss-constrained lightweight convolutional neural networks for cervical cell classification. *Sensors* 2022; 22: 3272.
46. Devarajan D, Alex DS, Mahesh T, et al. Cervical cancer diagnosis using intelligent living behavior of artificial jellyfish optimized with artificial neural network. *IEEE Access* 2022; 10: 126957–126968.
47. Fang M, Fu M, Liao B, et al. Deep integrated fusion of local and global features for cervical cell classification. *Comput Biol Med* 2024; 171: 108153.
48. Kalbhor M, Shinde S, Popescu DE, et al. Hybridization of deep learning pre-trained models with machine learning classifiers and fuzzy min–max neural network for cervical cancer diagnosis. *Diagnostics* 2023; 13: 1363.
49. Attallah O. CerCan· Net: cervical cancer classification model via multi-layer feature ensembles of lightweight CNNs and transfer learning. *Expert Syst Appl* 2023; 229: 120624.
50. Chen H, Liu J, Jiang P, et al. MSCCNet: multi-scale convolution-capsule network for cervical cell classification. In: *2023 IEEE international conference on bioinformatics and biomedicine (BIBM)*, pp.1842–1845. IEEE.
51. Pramanik R, Banerjee B and Sarkar R. MSENet: mean and standard deviation based ensemble network for cervical cancer detection. *Eng Appl Artif Intell* 2023; 123: 106336.
52. Maurya R, Pandey NN and Dutta MK. VisionCervix: Papanicolaou cervical smears classification using novel CNN-vision ensemble approach. *Biomed Signal Process Control* 2023; 79: 104156.
53. Yaman O and Tuncer T. Exemplar pyramid deep feature extraction based cervical cancer image classification model using Pap-smear images. *Biomed Signal Process Control* 2022; 73: 103428.
54. Mousser W, Ouadfel S, Taleb-Ahmed A, et al. IDT: an incremental deep tree framework for biological image classification. *Artif Intell Med* 2022; 134: 102392.